# A new active disturbance controller based on an improved fraction-order extended state observer

Bolin Li
*School of Artificial Intelligence and Automation, Huazhong University of Science and Technology*
Wuhan, China
libolin5043@foxmail.com

Lijun Zhu*
*School of Artificial Intelligence and Automation, Huazhong University of Science and Technology*
Wuhan, China
ljzhu@hust.edu.cn

*Abstract*—**This paper proposes a new fraction-order active disturbance rejection controller based on an improved fraction-order extended state observer (IFESO) for a class of fraction-order systems. Applying the IFESO, the fraction-order system can be approximated as an integer-order integrator ($1/s$). The parameters that ensure the stability of the IFESO and closed-loop system are provided. The frequency-domain analysis shows that the IFADRC has a stronger disturbance estimation performance for the fraction-order system than an integer-order active disturbance rejection controller (IADRC). The simulation results demonstrates that the proposed IFADRC behaves better than the IADRC and the FADRC.**

## I. INTRODUCTION

Fractional calculus has been developed for about 300 years and applied to different industrial fields in recent years [1]. With the in-depth understanding of the system, there are increasing demands to apply fractional calculus to control and modelling [2]. Many phenomena demonstrate fraction-order characteristics, such as fractal, and damping, to name a few [3-4]. The properties between resistance and capacitance can be modelled by fractional calculus [5-6]. In bioengineering, many systems are modelled or fitted as fraction-order systems [7-8]. A fraction-order modelling method was proposed in [9] for permanent magnet synchronous motor. Due to the introduction of an additional non-integer order, the fraction-order controller has the potential to achieve better and more robust control performance [10]. So far, many fraction-order controller counterparts of classic integer-order controller have been proposed, such as fraction-order sliding mode controller [11], fraction-order intelligent PID controller [12], fraction-order PID controller [14], and so on.

The core idea of active disturbance rejection control proposed by Han [15] is to improve the robustness of the system by using extended state observer (ESO) to estimate the internal and external disturbances in the system. The nonlinear structure of ADRC limits its application in the engineering field. A linear active disturbance rejection controller (ADRC) was proposed by Gao [16] to simplify the structure and has been widely applied. Many research results show that fraction-order active disturbance rejection controller (FOADRC) outperforms integer-order ADRC in terms of robustness and disturbance rejection performance [17-18]. A FOADRC control structure consisting of a proportional controller and a fraction-order ESO was proposed by Chen [19], demonstrating obvious advantages in terms of the robustness against high-frequency noise and disturbance over integer-order ADRC. An ADRC and fraction-order PID (FOPID) hybrid control scheme for the hydro-turbine speed governor system was proposed in [20] and is robust to the load disturbance.

ESO, an important part of ADRC, is used to estimate the total disturbance in the system. A simple integer-order control scheme for the fraction-order system based on an active disturbance rejection method was proposed in [21], which adopted an integer-order ESO (IESO). The IESO has poor estimation performance for high-frequency disturbances in fraction-order systems. A fraction-order active disturbance rejection control scheme based on a fraction-order ESO (FESO) was proposed by Li [22] to improve the performance of fraction-order systems. A fraction-order system can be converted to a fraction-order integrator by using the FESO.

In this paper, an IFESO is proposed to estimate the internal and external disturbances of a fraction-order system. The IFESO can convert a fraction-order system into an integer-order integrator. The ADRC based on the proposed IFESO is more robust than that based on the IESO. The main contributions of this paper is three-fold. First, the IFESO is proposed to convert a fraction-order system into an integer-order integrator and together with a proportional controller constitutes the improved fraction-order active disturbance rejection controller (IFADRC). Secondly, the stability criteria of IFESO and closed-loop control systems consisting of an fraction-order plant, and IFADRC are proposed. Third, it will be shown that IFADRC has advantages in terms of robustness and dynamic response performance compared with the IADRC and the FADRC. This paper is organized as follows. The IFADRC control structure is proposed in Section Ⅱ which is compared with that of IADRC and FADRC. The stability of the IFESO and the IFADRC closed-loop control system is analyzed in Section Ⅲ. The frequency-domain analysis of the three active disturbance rejection control structures are given in Section Ⅳ, and Section Ⅴ presents the simulation results.

## Ⅱ. Structure of Active Disturbance controls

The traditional ADRC consists of a nonlinear tracking differentiator, an extended state observer, and a nonlinear feedback control law. By the efforts of Gao [16], the bandwidth parameterization method was proposed to linearize the ESO and P controller without losing efficiency [23]. The IADRC was proved by Li [21] to estimate the disturbances in fraction-order systems, which considers the fraction-order dynamics as a common disturbance to present an integer-order system that is a single integer-order integrator. A fraction-order active disturbance rejection controller (FADRC) was proposed by Li [22] to convert the fraction-order system to a fraction-order integrator, which is a new control strategy for fraction-order systems. In this section, the structure of the IFADRC was presented. Compared with the IADRC, the IESO is replaced by an IFESO with an additional output signal. Unlike the FADRC, the IFADRC is used to convert the fraction-order system into an integer-order integrator.

Consider the single-input-single-output fraction-order control system, whose transfer function is described as follows:

$$G_f(s) = \frac{b_o}{s^\mu + a_o} \qquad (1)$$

where $a_o$, $b_o$, $\mu$ ($0 < \mu < 1$) are constants, $s$ is a Laplace operator. Note that the system parameters $a_o$ and $b_o$ might not be known. The state-space representation of the fraction-order system in (1) with the external disturbance, denoted by $\omega$ as follows:

$$y^{(\mu)} + a_o y = b_o u + \omega \qquad (2)$$

where $u$ and $y$ are the input and output of the fraction-order system $G_f(s)$, respectively. Define the quantity $q = \dot{y} - y^{(\mu)}$. Equation (2) can be rewritten as follows:

$$\dot{y} = \dot{y} - y^{(\mu)} - a_0 y + (b_0 - b)u + bu + \omega = f_{ifo} + bu + q \quad (3)$$

where $f_{ifo} = -a_o y + (b_0 - b)u + \omega$ is called the total disturbance. In particular, the term $-a_o y + (b_o - b)u$ is caused by the unknown internal dynamic and thus called internal disturbance, while the term $\omega$ is the external disturbance.

Define $h_{ifo} = \dot{f}_{ifo}$, $x_1 = y$ and $x_2 = f_{ifo}$. The state-space representation of (3) is given as follows:

$$\begin{pmatrix} \dot{x}_1 \\ \dot{x}_2 \end{pmatrix} = \begin{pmatrix} 0 & 1 \\ 0 & 0 \end{pmatrix}\begin{pmatrix} x_1 \\ x_2 \end{pmatrix} + \begin{pmatrix} b \\ 0 \end{pmatrix}u + \begin{pmatrix} 1 \\ 0 \end{pmatrix}q + \begin{pmatrix} 0 \\ 1 \end{pmatrix}h_{ifo} \quad (4)$$

where $x_2$ is an extended state. The improved fraction-order ESO (IFESO) is thus proposed as follows:

$$\begin{cases} \begin{pmatrix} \dot{z}_1 \\ \dot{z}_2 \end{pmatrix} = \begin{pmatrix} 0 & 1 \\ 0 & 0 \end{pmatrix}\begin{pmatrix} z_1 \\ z_2 \end{pmatrix} + \begin{pmatrix} b \\ 0 \end{pmatrix}u + \begin{pmatrix} 1 \\ 0 \end{pmatrix}\hat{q} + L(y - \hat{y}) \\ \hat{q} = (\dot{z}_1 - z_1^{(\mu)}) \\ \hat{y} = \begin{pmatrix} 1 & 0 \end{pmatrix}\begin{pmatrix} z_1 \\ z_2 \end{pmatrix} \end{cases} \quad (5)$$

where $\hat{q}$ and $\hat{y}$ are the outputs of IFESO, $z_1$ and $z_2$ are the estimation of $x_1$ and $x_2$, respectively, as well as $L = \begin{pmatrix} \beta_1 & \beta_2 \end{pmatrix}^T$ is the observer gain. The structure of the IFADRC is shown in Fig. 1.

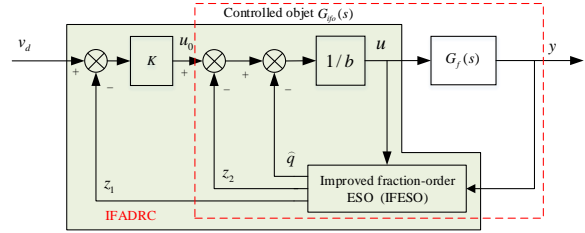

Fig. 1. Structure of the IFADRC

Similar to [21], the design principle of IFESO is to render the controlled object $G_{ifo}(s)$ as an integer-order integrator. Note that the relationship between the $u_0$ and $u$ is as follows:

$$\frac{u_0 - \hat{f}_{ifo} - \hat{q}}{b} = u \qquad (6)$$

Substituting (6) into (3) gives:

$$\dot{y} = f_{ifo} - \hat{f}_{ifo} + q - \hat{q} + u_0 \qquad (7)$$

When $f_{ifo}$ and $q$ are well estimated by $\hat{f}_{ifo}$ and $\hat{q}$, then the system $G_{ifo}(s)$ becomes an integer-order integrator $\dot{y} = u_o$. The stability of the IFESO is given in Section Ⅲ and the detailed frequency-domain analysis is given in Section Ⅳ.

For purpose of comparison, the IADRC and FADRC are also reviewed as follows. The major differences among IFADRC, IADRC and FADRC lie in the selection of the extended state to be estimated, while the structures of the extended state estimators, named IFESO, IESO and FESO, respectively, are more or less very similar.

The integer-order ESO (IESO) is given:

$$\begin{cases} \begin{pmatrix} \dot{z}_1 \\ \dot{z}_2 \end{pmatrix} = \begin{pmatrix} 0 & 1 \\ 0 & 0 \end{pmatrix} \begin{pmatrix} z_1 \\ z_2 \end{pmatrix} + \begin{pmatrix} b \\ 0 \end{pmatrix} u + L(y - \hat{y}) \\ \hat{y} = \begin{pmatrix} 1 & 0 \end{pmatrix} \begin{pmatrix} z_1 \\ z_2 \end{pmatrix} \end{cases} \qquad (8)$$

where $\hat{y}$ is the output of IESO, $z_1$ and $z_2$ are the estimator states. In particular, $z_1$ is the estimator for $x_1 = y$ and $z_2$ is the estimator for the extended state $x_2 = f_{io}$ where $f_{io} = \dot{y} - y^\mu - a_o y + (b_o - b)u + \omega$. Define $h_{io} = \dot{f}_{io}$. The state-space equation form of (2) is as follows:

$$\begin{pmatrix} \dot{x}_1 \\ \dot{x}_2 \end{pmatrix} = \begin{pmatrix} 0 & 1 \\ 0 & 0 \end{pmatrix} \begin{pmatrix} x_1 \\ x_2 \end{pmatrix} + \begin{pmatrix} b \\ 0 \end{pmatrix} u + \begin{pmatrix} 0 \\ 1 \end{pmatrix} h_{io} \qquad (9)$$

The structure of the IADRC is shown in Fig. 2.

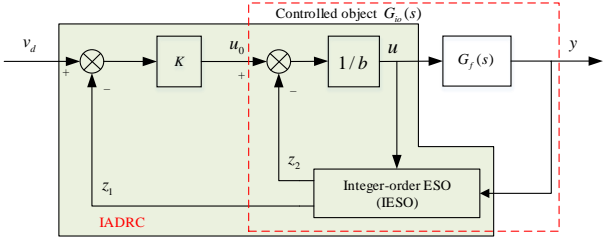

Fig. 2.   Structure of the IADRC

The fraction-order ESO (FESO) is given:

$$\begin{cases} \begin{pmatrix} z_1^{(\mu)} \\ z_2^{(\mu)} \end{pmatrix} = \begin{pmatrix} 0 & 1 \\ 0 & 0 \end{pmatrix} \begin{pmatrix} z_1 \\ z_2 \end{pmatrix} + \begin{pmatrix} b \\ 0 \end{pmatrix} u + L(y - \hat{y}) \\ \hat{y} = \begin{pmatrix} 1 & 0 \end{pmatrix} \begin{pmatrix} z_1 \\ z_2 \end{pmatrix} \end{cases} \qquad (10)$$

where $\hat{y}$ is the output of FESO, $z_1$ and $z_2$ are the estimator states. In particular, $z_1$ is the estimator for $x_1 = y$ and $z_2$ is the estimator for the extended state $x_2 = f_{fo}$ where $f_{fo} = -a_o y + (b_o - b)u + \omega$. Define $h_{fo} = f_{fo}^{(\mu)}$. The state-space equation form of (2) is as follows:

$$\begin{pmatrix} x_1^{(\mu)} \\ x_2^{(\mu)} \end{pmatrix} = \begin{pmatrix} 0 & 1 \\ 0 & 0 \end{pmatrix} \begin{pmatrix} x_1 \\ x_2 \end{pmatrix} + \begin{pmatrix} b \\ 0 \end{pmatrix} u + \begin{pmatrix} 0 \\ 1 \end{pmatrix} h_{fo} \qquad (11)$$

The structure of the FADRC is shown in Fig. 3.

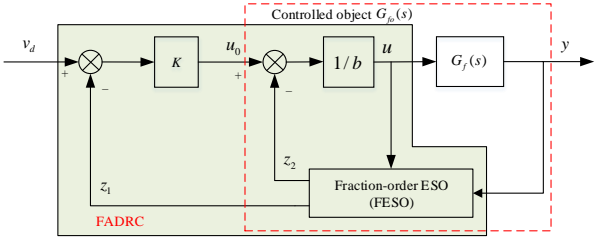

Fig. 3.   Structure of the FADRC

## III.   STABILITY ANALYSIS

### A.   Stability of the IFESO

Let $\dot{e}_i = \dot{x}_i - \dot{z}_i, i = 1, 2$, from (4) and (5), the error equation is as follows:

$$\begin{pmatrix} \dot{e}_1 \\ \dot{e}_2 \end{pmatrix} = \begin{pmatrix} -\beta_1 & 1 \\ -\beta_2 & 0 \end{pmatrix} \begin{pmatrix} e_1 \\ e_2 \end{pmatrix} + \begin{pmatrix} \dot{e}_1 - e_1^{(\mu)} \\ 0 \end{pmatrix} + \begin{pmatrix} 0 \\ 1 \end{pmatrix} h_{ifo} \qquad (12)$$

From (16), the error equation can be obtained as follows:

$$\begin{pmatrix} e_1^{(\mu)} \\ \dot{e}_2 \end{pmatrix} = \begin{pmatrix} -\beta_1 & 1 \\ -\beta_2 & 0 \end{pmatrix} \begin{pmatrix} e_1 \\ e_2 \end{pmatrix} + \begin{pmatrix} 0 \\ 1 \end{pmatrix} h_{ifo} \qquad (13)$$

**Theorem 1.** If $\beta_1 > 0$ and $\beta_2 > 0$, $\forall \mu \in (0,1)$, the IFESO is bounded-input bounded-output (BIBO) stable, regarding $u$, $y$ as the input and $z_1$, $z_2$ as the output.

**Proof.** The proof is omitted due to space limitation..

**Remark.** For fair comparison and simplified analysis, the frequency-domain analysis and simulation illustration in this paper ensure that $\beta_1 = 2\omega_0$, $\beta_2 = \omega_0^2$.

### B.   Stability analysis for the closed-loop system

**Theorem 2.** Let $p$, $q$ be position integers satisfying $p/q = \mu$, $\lambda = 1/q$, $\beta_1 = 2\omega_o$ and $\beta_2 = \omega_o^2$ in IFESO (5), where $\omega_o$ is some positive value. If the parameters $K$ of the proportional controller guarantees the following formula holds:

$$|\arg(w_i)| > \lambda \frac{\pi}{2} \qquad (14)$$

where $w_i$ is the ith root of (15):

$$\begin{aligned} & bw^{2q+p} + (b_o \beta_1 + a_o b)w^{2q} + (bK + \beta_1 b - \beta_1 b_o)w^{q+p} \\ & + (a_o b \beta_1 + a_o bK + Kb_o \beta_1 + b_o \beta_2)w^q + b_o \beta_2 K = 0 \end{aligned} \qquad (15)$$

the closed-loop control system is BIBO stable, regarding $v_d$ as the input and $y$ as the output..

**Proof.** The proof is omitted due to space limitation.

## IV.   FREQUENCY-DOMAIN ANALYSIS

Based on the fraction-order control system (1) in Section II, ignoring the influence of external disturbance, i.e, $\omega = 0$, the frequency-domain analysis for the proposed IFADRC, and the IADRC are presented in this section.

### A.   Frequency-domain analysis of IADRC

Conducting Laplace transform to the IESO gives:

$$\begin{cases} Z_1(s) = \dfrac{2\omega_o s + \omega_o^2}{s^2 + 2\omega_o s + \omega_o^2} Y(s) + \dfrac{bs}{s^2 + 2\omega_o s + \omega_o^2} U(s) \\ Z_2(s) = \dfrac{\omega_o^2 s}{s^2 + 2\omega_o s + \omega_o^2} Y(s) - \dfrac{\omega_o^2 b}{s^2 + 2\omega_o s + \omega_o^2} U(s) \end{cases} \qquad (16)$$

where $Z_1(s)$, $Z_2(s)$, $Y(s)$, and $U(s)$ are the Laplace transforms corresponding to $z_1$, $z_2$, $y$, and $u$, respectively.

From equations (1) and (16), the transfer function of the controlled object $G_{io}(s)$ can be obtained as follows:

$$G_{io}(s) = \frac{U_o(s)}{Y(s)} = \frac{b_o s + \omega_o{}^2}{s(b_o \omega_o{}^2 + a_o b(s + 2\omega_o) + b s^\mu (s + 2\omega_o))} \quad (17)$$

According to equation (17), the transfer function of the controlled object $G_{io}(s)$ is related to the controlled object $G_f(s)$ and $\omega_o$.

### B. Frequency-domain analysis of IFADRC

Conducting Laplace transform on the IFESO gives:

$$\begin{cases} Z_1(s) = \dfrac{2\omega_o s + \omega_o{}^2}{s^{\mu+1} + 2\omega_o s + \omega_o{}^2} Y(s) + \dfrac{bs}{s^{\mu+1} + 2\omega_o s + \omega_o{}^2} U(s) \\[2mm] Z_2(s) = \dfrac{\omega_o{}^2 s^\alpha}{s^{\mu+1} + 2\omega_o s + \omega_o{}^2} Y(s) - \dfrac{b\omega_o{}^2}{s^{\mu+1} + 2\omega_o s + \omega_o{}^2} U(s) \\[2mm] Q(s) = (s - s^\mu) Z_1(s) \end{cases}$$

$$(18)$$

where $Z_1(s)$, $Z_2(s)$, $Y(s)$, $U(s)$ and $Q(s)$ are the Laplace transforms corresponding to $z_1$, $z_2$, $y$, and $u$, and the quantity $q$, respectively.

From equations (1) and (18), the transfer function of the controlled object $G_{ifo}(s)$ can be obtained as follows:

$$G_{ifo}(s) = \frac{U_o(s)}{Y(s)} = \frac{b_o(s^{1+\mu} + 2\omega_o s + \omega_o{}^2)}{s(b_o \omega_o(2s - 2s^\mu + \omega_o) + a_o b(s + 2\omega_o) + b s^\mu (s + 2\omega_o))}$$

$$(19)$$

According to (19), the transfer function of the controlled object $G_{ifo}(s)$ is related to the controlled object $G_f(s)$ and $\omega_o$.

### C. Performance analysis of disturbance estimation

Both the IADRC and the IFADRC are meant to match the compensated object $G_o(s)$ ($G_{io}(s)$ or $G_{ifo}(s)$) consisting of the plant and extended state estimator to an integer-order integrator. In order to compare the performance of the IADRC and the IFADRC, define

$$\Delta_o = 1 - j\omega G_o(j\omega) \quad (20)$$

$\Delta_o$ can be regarded as the difference between an integer-order single integrator and the compensated object $G_o(s)$ at frequency $\omega$. In [27], the mean-square error (MSE)

$$e_o = |\Delta_o|^2 \quad (21)$$

is used to evaluate the mismatch between two models. In what follows, the mean-square error in (21) is also used. Moreover, $b = b_o$ is imposed for the following analysis.

From (17) and (21), the mean-square error of the frequency value between an integer-order single integrator and $G_{io}(s)$ can be expressed as follows:

$$e_{io} = \frac{N_1}{D_1} \quad (22)$$

where

$$\begin{cases} N_1 = (\beta_1{}^2 + \omega^2)(a_o{}^2 + \omega^2 + \omega^{2\mu} - 2\omega^\mu(\omega\cos(\tfrac{\pi\mu}{2}) - ao\sin(\tfrac{\pi\mu}{2}))) \\[2mm] D_1 = (a_o\beta_1 + \beta_2 + \omega^\mu(\omega\cos(\tfrac{\pi\mu}{2}) + a_o\sin(\tfrac{\pi\mu}{2})))^2 \\[2mm] \quad + (2\beta_1\omega + \omega^\mu(a_o\cos(\tfrac{\pi\mu}{2}) - \omega\sin(\tfrac{\pi\mu}{2})))^2 \\[2mm] \beta_1 = 2\omega_o \\[2mm] \beta_2 = \omega_o{}^2 \end{cases}$$

$$(23)$$

From (19) and (21), the mean-square error of the frequency value between an integer-order single integrator and $G_{ifo}(s)$ can be expressed as follows:

$$e_{ifo} = \frac{N_2}{D_2} \quad (24)$$

where

$$\begin{cases} N_2 = a_o{}^2(\beta_1{}^2 + \omega^2) \\[2mm] D_2 = (a_o\beta_1 + \beta_2 + \omega^{1+\mu}\cos(\tfrac{\pi\mu}{2}))^2 \\[2mm] \quad + \omega^2(a_o + 2\beta_1 - \omega^\mu\sin(\tfrac{\pi\mu}{2}))^2 \\[2mm] \beta_1 = 2\omega_o \\[2mm] \beta_2 = \omega_o{}^2 \end{cases}$$

$$(25)$$

According to the principles of the IESO and the IFESO, the controlled object ( $G_{io}(s)$ or $G_{ifo}(s)$ ) should be approximately equivalent to an integer-order integrator. Fig. 4 is the curves of the mean-square error $e_{io}$ and the mean-square error $e_{ifo}$. It is shown in Fig. 4 that the IFADRC can better convert the fraction-order control system into an integer-order integrator than the IADRC. Fig. 5 shows the curves of the mean-square error $e_{io}$ and the mean-square error $e_{ifo}$ under different $\mu$. Fig. 6 shows the curves of the mean-square error $e_{io}$ and the mean-square error $e_{ifo}$ under different $a_o$. It is shown in Fig. 7 that the variation of the mean-square error with $\omega_o$. As shown in Fig. 5, Fig. 6, and Fig. 7, the variation of the mean-square error $e_{ifo}$ with $\mu$, $a_o$, and $\omega_o$ is smaller than that of the mean-square error $e_{io}$, indicating that the IFADRC control system is more robust to $\mu$, $a_o$, and $\omega_o$ variation than the IADRC control system.

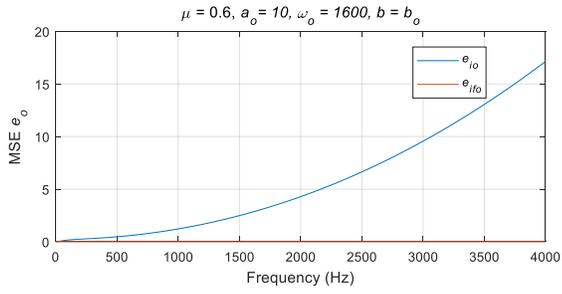

Fig. 4. The MSE curves when $a_o = 10$, $\omega_o = 1600$, and $\mu = 0.6$

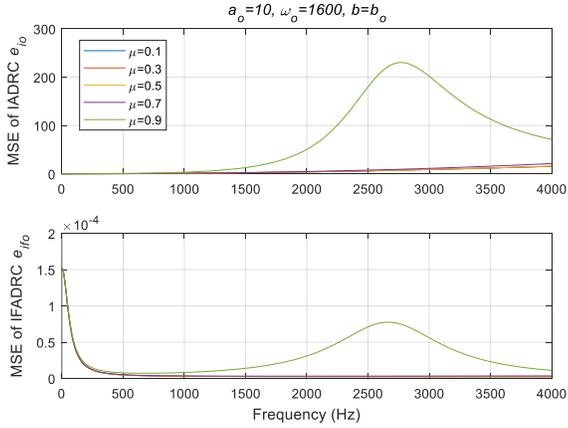

Fig. 5. The MSE curves under different $\mu$ when $a_o = 10$ and $\omega_o = 1600$

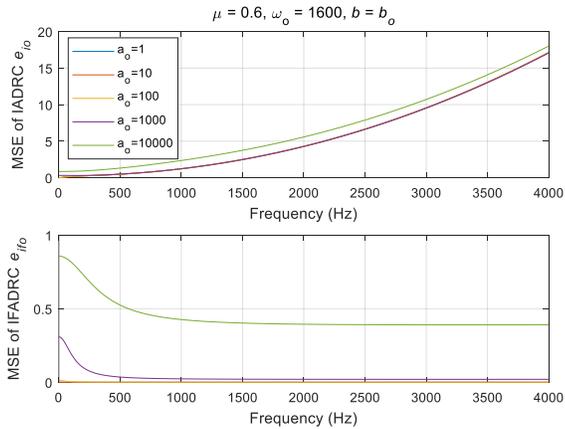

Fig. 6. The MSE curves under different $a_o$ when $\mu = 0.6$ and $\omega_o = 1600$

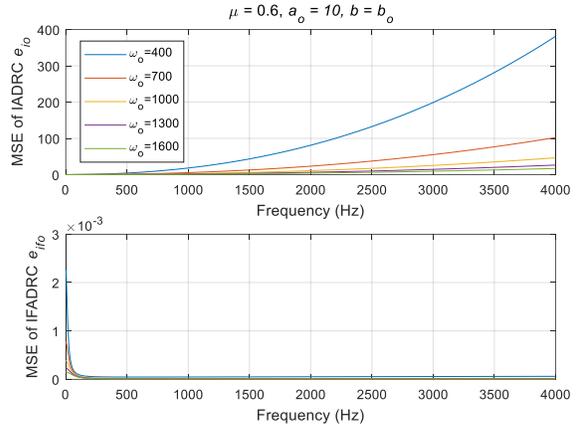

Fig. 7. The MSE curves under different $\omega_o$ when $\mu = 0.6$ and $a_o = 10$

The comparison results of the Bode diagram of the controlled object $G_{io}(s)$ and the Bode diagram of the controlled object $G_{fo}(s)$ are shown in Fig. 8 and Fig. 9. It is shown in Fig. 8 and Fig. 9 that the frequency characteristic of $G_{fo}(s)$ can be fitted equivalent to the frequency characteristic of an integer-order integrator in the whole frequency band, but the frequency characteristic of $G_{io}(s)$ cannot. When the order of the controlled object $G_f(s)$ increases, the estimation effect of IESO in the high-frequency band varies greatly, but the estimation effect of IFESO in the high-frequency band remains unchanged.

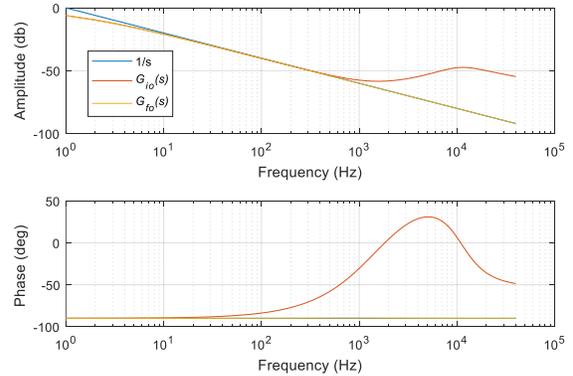

Fig. 8. Bode diagram of the controlled object $G_{io}(s)$ ( $\mu = 0.6$ )

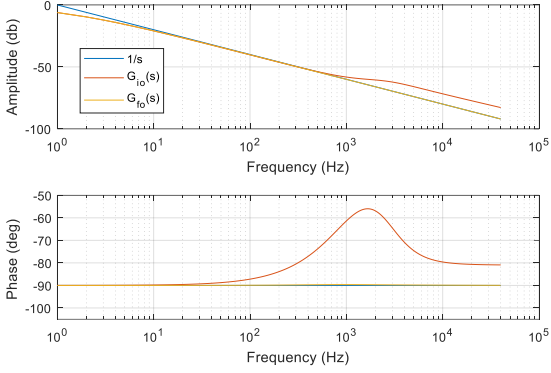

Fig. 9. Bode diagram of the controlled object $G_{fo}(s)$ ($\mu = 0.9$)

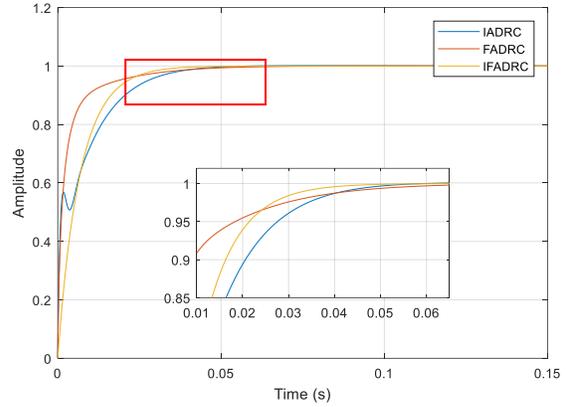

Fig. 11. Step response of different control methods

## V. SIMULATION ILLUSTRATION

In order to verify the control effect of the IFADRC, the simulation results of three different control methods using Matlab/Simulink are given in this section. The transfer function of the controlled object $G_f(s)$ is considered as follows:

$$G_f(s) = \frac{1}{s^{0.8} + 10} \qquad (26)$$

For a fair comparison, $K$ and $\omega_o$ are equal for different control methods. The simulation is compared when $b$ is equal to $b_o$. The discrete frequency of IFESO is 8000Hz, and the discrete order is 5. Set $K=150$, $\beta_1 = 2\omega_o$, $\beta_2 = {\omega_o}^2$, and $\omega_o = 400$ rad/s.

According to Theorem 1, $\beta_1 = 2\omega_o$ and $\beta_2 = {\omega_o}^2$, the IESO is BIBO stable. From (26), $\mu = 0.8$, $a_o = 10$ and $b_o = 1$. Refer to section V, $p = 4$, $q = 5$, and $\lambda = 0.2$.

Substitute the parameters of the control system into (15), $w_i$ can be solved. All $w_i$ can be got as shown in Fig. 10 and are in stable areas, thus the IFADRC closed-loop control system in the subsection is BIBO stable.

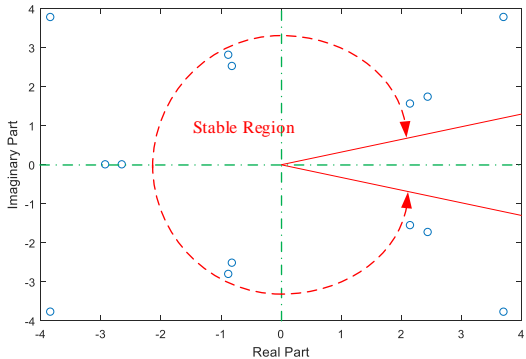

Fig. 10. Root locus of the IFADRC closed-loop control system

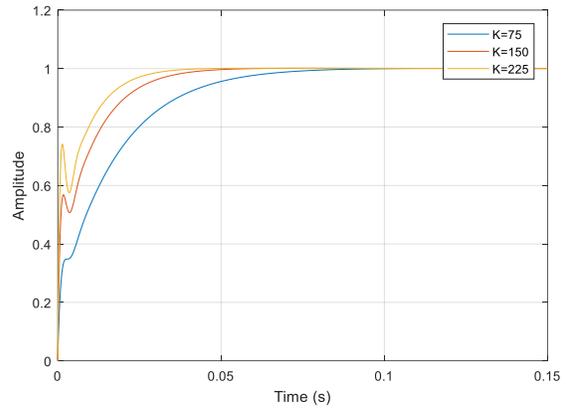

Fig. 12. Step response of the IADRC with loop gain variation

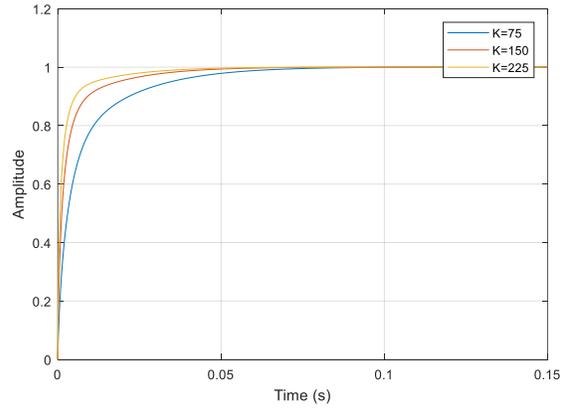

Fig. 13. Step response of the FADRC with loop gain variation

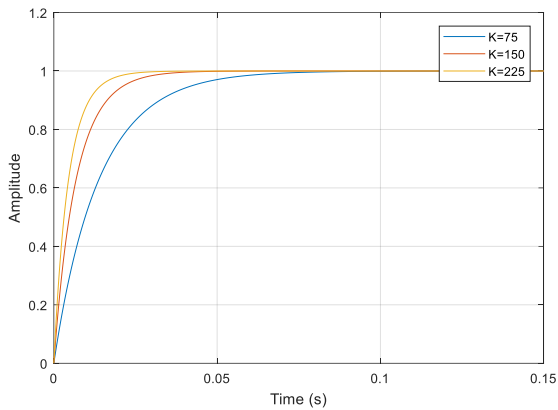

Fig. 14. Step response of the IFADRC with loop gain variation

The step responses of the IADRC control system, the FADRC control system, and the IFADRC control system are shown in Fig. 11. It is shown in Fig. 11 that the IFADRC control system has better dynamic response performance than the IADRC control system and the FADRC control system. Fig. 12, Fig. 13 and Fig. 14 are the step response of different control systems with loop gain variation. As shown from Fig. 12, Fig. 13 and Fig. 14, when $b$ equals $b_o$, the FADRC control system and the IFADRC control system are robust to loop gain variation.

## VI. Conclusions

In this paper, a new active disturbance structure based on an improved ESO is proposed. The fraction-order system can be converted into a unit integer-order integrator. In addition, this paper also presents the stability criterion of the IFADRC control system. The IFADRC is robust to loop gain variation. Simulation results are illustrated that IFESO has better disturbance estimation performance than IESO and the IFADRC outperforms the IADRC and the FADRC.

**Authors' background**

| Name | Prefix | Research Field | Email | Personal website |
|---|---|---|---|---|
| Bolin Li | Master Student | Fraction-order control | libolin5043@foxmail.com | |
| Lijun Zhu | Full Professor | 1.Flexible, bionic robot system modeling and control  2.Networked analysis and control | ljzhu@hust.edu.cn | |
| | | | | |
| | | | | |
| | | | | |

**Note:**

**[1] This form helps us to understand your paper better; <span style="color:red">the form itself will not be published.</span>**

**[2] *Prefix*:** can be chosen from Master Student, PhD Candidate, Assistant Professor, Lecture, Senior Lecture, Associate Professor, Full Professor